\documentclass[12pt,preprint]{aastex}

\newcommand{\etal}{\mbox{\it{et al.}~~}}
\newcommand{\nd}{...}
\begin{document}
%%%%%%%%%%%%%    TITLE     %%%%%%%%%%%%%%%

\title{Infrared spectroscopy of nearby radio active elliptical galaxies}% that are radiosources}
%%%%%%%%%%%%   AUTHORS     %%%%%%%%%%%%%%
%\author{\parbox{\textwidth}{\flushleft
%\vspace{-0.5cm}
%
% Please indicate only one corresponding author email, as per the following example:
%{\it
\author{Jeremy Mould$^{1,2}$, Tristan Reynolds$^{3}$, Tony Readhead$^{4}$,David Floyd$^{5}$, Buell Jannuzi$^{6}$, Garret Cotter$^8$, Laura Ferrarese$^{7}$, Keith Matthews$^{4}$, David Atlee$^{6}$, Michael Brown$^{5}$}%}\\

\vspace{0.4cm}
%
%{\small 
% \affil{$^{1}$Swinburne University,\ Hawthorn Vic 3122}\\
% \affil{$^{2}$ARC Centre of Excellence for All-sky Astrophysics (CAASTRO)},\\
% \affil{$^{3}$University of Melbourne},\\
% \affil{$^{4}$California Institute of Technology},\\
% \affil{$^{5}$Monash University},\\
% \affil{$^{6}$Steward Observatory, University of Arizona (formerly at NOAO)},\\
% \affil{$^{7}$Herzberg Institute of Astrophysics} \\
% \affil{$^{8}$Department of Astrophysics, Oxford University} 
 \affil{Swinburne University$^1$, Hawthorn Vic 3122, Australia}
 \affil{ARC Centre of Excellence$^2$ for All-sky Astrophysics (CAASTRO)}
 \affil{University of Melbourne$^3$}
 \affil{California Institute of Technology$^4$}
 \affil{School of Physics, Monash University$^5$}
 \affil{Steward Observatory, University of Arizona$^6$ (formerly at NOAO)}
 \affil{Herzberg Institute of Astrophysics$^7$} 
 \affil{Department of Physics, University of Oxford, Denys$^8$}
%{\small 
%My formal affiliation/address is:
%Wilkinson Building, Keble Road, Oxford OX1 3RH, United Kingdom
\authoraddr{Email: jmould@swin.edu.au}
%}}}
%
%
%%%%%%%%        DO NOT EDIT FOLLOWING     %%%%%%%%%%%%
%\begin{document}

%\twocolumn[
%\begin{changemargin}{.8cm}{.5cm}
%\begin{minipage}{.9\textwidth}
%\vspace{-1cm}
%\maketitle
%
%
%%%%%%%%%%%%%     ABSTRACT    %%%%%%%%%%%%%

\small{\bf Abstract: }
In preparation for a study of their circumnuclear gas we have surveyed 60\% of a complete sample of elliptical galaxies within 75 Mpc that are radiosources.
Some 20\% of our nuclear spectra have infrared emission lines, mostly Paschen lines, Brackett$\gamma$ and [FeII]. We consider the influence of radio power and
black hole mass in relation to the spectra. Access to the spectra is provided
here as a community resource.

%%%%%%%%%%%%%     KEYWORDS    %%%%%%%%%%%%%
\medskip{\bf Keywords:} infrared: general --- active galactic nuclei -- galaxies: elliptical -- radiosources
% standard list of subject headings adopted by The Astrophysical Journal
% and available from http://www.journals.uchicago.edu/ApJ/keywords_text.html.

%%%%%%%%DO NOT EDIT%%%%%%%%%%%%
\medskip
\medskip
%\end{minipage}
%\end{changemargin}
%]
%\small
%%%%%%%%EDIT FROM HERE%%%%%%%%%%%%

\section{Introduction}

Astrophysicists have striven to learn the detailed physics of Active Galactic Nuclei (AGN) since black holes were first posited as the energy source
(Lynden-Bell \& Rees 1971). Landmarks were the dusty torus model (Mayes, Pearce \& Evans 1985) 
and the unified scheme of AGN which realised the implications of this geometry (Antonucci \etal 1987) from the observer's standpoint.
With the advent of Laser Guide Star Adaptive Optics on 10 metre class telescopes, the opportunity now exists to image the circumnuclear gas in nearby
active galaxies and, with the aid of detailed modelling, to understand the physics of the central 10-20 light years of these powerful engines.
The purpose of the project our small team has recently commenced is to study a complete sample of galaxies that are radiosources within 40 Mpc.
Our work focuses first on the elliptical galaxies, as these are the simplest in terms of stellar population and gas and dust content.
We emphasize nearby galaxies, as at 20 Mpc the diffraction limited resolution of the OSIRIS instrument of the Keck Observatory is 3.2 pc,
which is appropriate for disk gas heated by AGN\footnote{Our sample mean is actually 40 Mpc, where the OSIRIS resolution is 6.4 pc.}.

Central black holes are now believed to be basic constituents of most, if not all, massive galaxies (Magorrian \etal 1998, Kormendy 2004). 
That black hole growth is
linked with galaxy formation was realised because of the correlation of
BH mass with the properties of the host galaxy (Kormendy 1993; Kormendy
\& Richstone 1995; Magorrian \etal 1998; Gebhardt \etal 2000; Ferrarese
\& Merritt 2000; Barth, Greene \& Ho 2005; Greene \& Ho 2006), although
alternative explanations have been put forth (Jahnke \& Maccio 2010).

The deviation of the galaxy mass function from the strict power-law describing dark matter halo masses provides an additional and independent argument in favor of feedback (e.g. Hopkins \etal 2008).  Feedback is also relevant to the cooling flow problem, particularly with respect to radio AGN (e.g. Bower \etal 2006, Best+06).
Studies have shown that $\sim$ 30\% of the most massive galaxies are radio continuum sources (e.g., Fabbiano \etal 1989; Sadler \etal 1989).
As a direct manifestation of accretion, and therefore black hole growth, AGN and the consequences of their energy feedback have figured prominently in most current ideas of structure formation (e.g., Granato \etal 2004; Springel, Di Matteo \& Hernquist 2005; Hopkins \etal 2006). Awareness of the importance of black holes has engendered broad interest in the study of the AGN phenomenon itself. 

This paper focuses on nuclear activity in nearby galaxies. By selection, most of the objects occupy the faintest end of the AGN luminosity function and have low accretion rates. Although energetically less impressive, low luminosity AGNs deserve scrutiny. %By virtue of the short duty cycle of black hole accretion (Greene \& Ho 2007), most AGNs spend their lives in a low state, such that the bulk of the population has relatively modest luminosities. 
By virtue of proposed reasons such as duty cycle (Greene and Ho 2007) or high/low accretions states (see e.g. Hardcastle, Evans \& Croston 2006), most AGNs are found in a low luminosity state. Brown \etal (2011) argue that all massive
early-type galaxies are radiosources, mostly powered by AGN (Slee \etal 1994;
Ho 1999).
The advent of new telescopes and new analysis techniques yield fresh insights into this problem. 

The aims of our project are (1) to image nearby ellipticals that are radiosources in 1--2$\mu m$ emission lines, (2) to deduce the kinematics of the gas in these lines,
(3) to model the density and temperature of the gas, (4) to deduce the mass of the nuclear black hole and the star formation rate in the central 10 pc,
(5) to measure gas inflow and outflow rates in non-equilibrium regions of the central structure and (6) summarize these results over the 40 Mpc complete sample of ellipticals, noting correlations with mass, radio power, star formation rate,
emission line strength and x-ray power.

In this paper we present the first half of a survey of infrared emission lines carried out at Palomar Observatory and Kitt Peak National Observatory. A number of the galaxies show lines of [FeII] which will be followed up with 8 meter class telescopes to accomplish the first goal of our project. Circumnuclear star formation rates are presented for those with Brackett $\gamma$ in emission. 

\section{Observations}
Brown \etal (2011) have compiled a complete sample to K = 9 of 396 elliptical galaxies with declination $>$ --40$^\circ$. The 231 detected radiosources\footnote{Galaxies for which upper limits were given in Brown's data table were omitted} among them are the basic sample for this paper.
For diffraction limited imaging of the centers of these galaxies with the Keck and Gemini Observatories we have begun to observe the subsample with
1.4 GHz radio power $>$ 10$^{21}$ W Hz$^{-1}$ and declination $>$ --36$^\circ$ with the instrumentation listed in Table 1.
Sample near infrared spectra are shown in Figure 1--4, in which no redshift correction has been applied. Currently, 20\% of the galaxies observed are showing emission lines of [FeII] and/or Br$\gamma$,
suitable for followup with Keck OSIRIS and Gemini North NIFS integral field unit (IFU) spectrographs\footnote{For all but the faintest nuclei, on-axis guiding is possible.}.
Our 4 runs (Table 1) with infrared slit spectrographs have netted nuclear spectra of 136 galaxies from the basic sample. 
This is sufficient to understand the statistics of circumnuclear properties in current epoch ellipticals. Areas of notable incompleteness are the greater Virgo cluster area and the southern hemisphere. Black hole masses in Table 4 are based on the K-band magnitude, following Graham (2007).

At both telescopes all galaxies were observed in two slit positions, A \& B, separated by 20$^{\prime\prime}$, in an ABBA sequence of 4--5 minute exposures.
Following the object sequence, an 8th magnitude A0V star was observed within an airmass of 0.1 of the object for telluric correction.
This was also used for spectrophotometric calibration which employed Vega model fluxes for zeroth visual magnitude. 
For wavelength calibration an argon lamp was used at Kitt Peak, telluric OH lines at Palomar.

The galaxies with emission lines in these near infrared spectra are listed in Table 2; those without are in Table~3, which also gives the signal to noise ratio of the Kitt Peak spectra at 1.01$\mu$ and Palomar spectra at 1.04$\mu$.

\section{Data Reduction}

%Spextool: A Spectral Extraction Package for SpeX, a 0.8-5.5 micron Cross-Dispersed Spectrograph

The data sets from each observatory required different data reduction packages.  The Kitt Peak National Observatory data was reduced using IRAF and the reduction of the Palomar Observatory data was done using Spextool v4.0 beta.
(Cushing, Vacca \& Rayner 2004).

The first step in the reduction of the Kitt Peak data was the creation of the flat field and dark images.  The flat field was created by subtracting the flat lamps-off image from the flat lamps-on image.  The first and second nights' data had separate flat field images.  The third night's data set did not contain flat field images, so the flat field image created for night 2 was also used for night 3 objects.
	
For the standard stars, from the observation files the dark image was subtracted and then the flat field image was divided out.  The spectrum was then extracted using the task `apall' in IRAF.  As the observation files for the galaxies were taken as ABBA, the galaxy spectra were extracted by creating  images for A-B and B-A files which were then divided by the flat field image.  The flat fielded A-B and B-A images were then rotated and cropped before running `apall' on each image.
	
The axes for the spectra were initially in terms of pixels and had to be wavelength calibrated using an argon arc spectrum.  This was achieved by picking five prominent peaks in the arc spectrum and identifying the pixel value corresponding to the wavelength of each peak.  These five values were then used to interpolate along the entire spectrum.
	
%The spectra for the Kitt Peak data have not been telluric corrected due to shifts between the galaxies and standard stars which create distortions in the final galaxy spectra in attempting to perform the telluric correction.
	
To create the flat field image used with the Palomar data, IRAF was first used to form a set of flat field images by subtracting the individual flat lamp-off images from the flat lamp-on images.  The flat field images created with IRAF were then used in the program `xspextool' in Spextool to create the master flat field.  The wavelength calibration was created in `xspextool' using any four galaxy images.  The spectra for the galaxies and standard stars were extracted in `xspextool' for A-B and B-A image sets using the master flat field image and the wavelength calibration.

The individual extracted spectra for each galaxy and standard star were then combined using `xcombspec' by selecting all images of each object and scaling all orders with respect to the spectrum with the greatest magnitude in the order 3 spectrum.
	
Telluric correction was carried out on the combined spectra for each galaxy using the combined standard star spectra with a similar air mass to the galaxy using the Spextool program `xtellcor'.  The hydrogen absorption lines in the standard star spectrum were rescaled for all orders and the telluric spectrum was constructed. The four orders (3,4,5,6) of the telluric corrected galaxy spectra were merged using `xmergeorders' to give a single spectrum.

Spectra are provided for readers desiring access to them\footnote{https://sites.google.com/site/radioactivegalaxies/}. Cesetti \etal (2009) show spectra of
NGC 4649, NGC 4697, IC 4296, NGC 4564, NGC 5077, and NGC 5576. Riffel \etal
(2006) show a spectrum of NGC 2110.

\section{Discussion}
We compare the radio power of our emission line galaxies with that of the full 397 galaxy
Brown \etal sample in Figure 6. Infrared emission line nuclei have 6 times
stronger radio emission than average at a given K band magnitude (or stellar mass). This does not stand out in the figure, but is a 4$\sigma$ effect, based on the line in Figure 6 and the scatter in the relation.
Both radio and infrared quantities here are from the data table of Brown \etal. The whole-sample correlation may be expected on general scaling grounds. A larger
galaxy hosts a larger black hole which in turn is responsible for more radio power. Indeed, Ho (2008) shows a strong correlation between 
radio loudness and x-ray luminosity divided by Eddington ratio. However, if our infrared emission line galaxies stand out, it may be because the BH in
these cases is currently being fed by the circumnuclear disk.

We also examined the thermal infrared fluxes of the emission line galaxies.
Only UGC 3426 and NGC 3516 had mid-infrared excesses in the AKARI Point Source
Catalog (Murakami \etal 2007): the former has a 9$\mu$/2.2$\mu$ flux ratio of 1.7 and the latter 1.0. The infrared flux of UGC 3426 rises a further factor of 6 at 18$\mu$. In general, the IRAS fluxes of our near-infrared emission line galaxies do not correlate well with the SFR in Table 2 (Figure 7).

The infrared emission line spectra in Table 2 are not all clones of one another.
We compare the Br$\gamma$ profiles of NGC 2787 and IC 630 in Figure 8. The former has a broad rounded profile, FWHM 750 km/sec. The latter has a sharp profile,
FWHM 200 km/sec. The former is a LINER (Balzano 1983), the latter a starburst nucleus in the Veron-Cetty \& Veron (2006) catalog. Both are morphologically classified
S0. These differences are a strong motivation to explore the spatial structure of the emitting gas at 10 meter class diffraction limited resolution.

In Figure 9 we show the relation between BH mass and SFR. The latter is calculated from Br$\gamma$ (Kennicutt 1998). SFR uncertainties were calculated
from the Br$\gamma$ uncertainties in Table 2.
We draw the uncertainty in the BH mass from Table 4 where both the (K , M$_{BH}$) and ($\sigma$, M$_{BH}$) relations are employed and the galaxies have secondary distance indicators.
It is largely a scatter plot, although the sample is modest, and there
is a hint of a correlation in the galaxy scaling sense. If the BH formation process were steady in solar masses/year, one might expect
a correlation. If the process is sporadic, dependent for example on fuelling events or circumnuclear disk instability, a correlation
would easily be masked. Monitoring Br$\gamma$ over time or high resolution analysis of any disk could shed light on this. Photoionization by hard radiation
from close to the BH could also strenghthen Br$\gamma$ over the value due
to star formation.

\section{Conclusions}

Some 20\% of ellipticals in the Brown \etal sample have emission line nuclei which will be followed up with IFU spectrographs on AO equipped telescopes.
The complete sample may have a somewhat lower emission fraction, as galaxies with optical emission lines (where this was known) tended to be observed first in order to provide immediate targets for IFU observing. Emission line nuclei are
accompanied by more powerful radio sources at a given galaxy mass.

The SFR in these nuclei has a median value of 0.4 M$_\odot$/yr and must be intermittent to attain the median black hole mass of 5 $\times$ 10$^8$ M$_\odot$ of these galaxies.

\acknowledgements

%\section*{Acknowledgments} 
Thanks go to Eilat Glickman for sharing her observing expertise at the Palomar 200 inch telescope and her copy of the Triplespec Spextool IDL program. This paper has also made use of IRAF which is distributed by NOAO. NOAO is operated by AURA under a cooperative agreement with NSF. The Caltech-Swinburne Collaborative Agreement allows the first author access to the Hale and Keck telescopes.
Survey astronomy is supported by the Australian Research Council through CAASTRO\footnote{www.caastro.org}.
The Centre for All-sky Astrophysics is an Australian Research Council Centre of Excellence, funded by grant CE11001020.
GC acknowledges support from STFC grants  ST/H002456/1 and ST/I003673/1. MB acknowledges the support from the Australian Research Council via Future Fellowship FT100100280 and Discovery Project DP110102174. We thank an anonymous referee
for comments which improved the paper.

%\begin{thebibliography}{}
% References are listed as in the following example, for more examples, please
% see the PASA Style Guide
%\bibitem[Ashley \etal (1996)]{mcba}
\noindent
Almudena Alonso-Herrero \etal 2000, ApJ 530 688\\
Antonucci, R. \etal 1987, AJ 93, 785\\
Balzano, V. 1983, ApJ, 268, 602\\
Barth, A., \etal 2001, ApJ, 555, 685\\
Barth A., Greene, J., \& Ho, L., 2005, ApJ, 619, L151\\
Best, P., Kaiser, C., Heckman, T., \& Kauffmann, G., 2006, MNRAS, 368, L67\\
Bower, R., Benson, A., Malbon, R., Helly, J., Frenk, C., Baugh, C., Cole, S., \& Lacey, C. 2006, MNRAS, 370, 645\\	
Brown, M.J.I., Jannuzi, B., Floyd, D., \& Mould, J. 2011, ApJ, 731, 41\\
Cesetti, M. \etal , 2009, A\&A, 497, 41\\%; Riffel+, 2006, A&A, 457, 61
Cushing, M., Vacca, W. \& Rayner, J. 2004, PASP, 116, 362\\
Fabbiano, G., Gioia, I. M., \& Trinchieri, G. 1989, ApJ, 347, 127\\
Ferrarese, L., \& Ford, H.F. 1999, ApJ, 515, 583\\
Ferrarese, L., \& Ford, H.F. 2005, Sp.Sc.Rev. 116, 523\\
Ferrarese L, \& Merritt D., 2000, ApJ, 539, L9 \\
Freedman, W.L., \etal, 2000, ApJ, 553, 47\\
Gebhardt, K., \etal 2007, ApJ, 671, 1321\\
Gebhardt K, Bender R, Bower G, Dressler A, Faber SM, \etal 2000, ApJ 539, L13\\
Granato GL, De Zotti G., Silva L., Bressan A., Danese L. 2004, ApJ, 600, 580\\
Graham, A. 2007, MNRAS, 379, 711\\
Green, R \& Schmidt, M. 1978, ApJ, 220, 1\\
Greene, JE, Ho, LC.2006, ApJ, 641, L21 \\
Greene, JE, Ho, LC.2007, ApJ, 667, 131 \\
Gultekin, K., \etal 2009, ApJ, 695, 1577\\
Hardcastle, M., Evans, D., \& Croston, J. 2006 MNRAS, 376, 1849\\
Ho, L. 1999, ApJ, 510, 631\\
Ho, L. 2008, ARAA, 46, 475\\
Hopkins, PF \& Hernquist, L. 2006, ApJ Suppl 166, 1 \\
Hopkins, PF \etal 2008, ApJS, 175, 356\\
Jahnke,K \& Maccio, A. 2010, ApJ, 734, 92\\
Jarrett, T.H. 2000, PASP, 112, 1008\\
Kennicutt, R 1998, ARAA, 36, 189\\
Kormendy J.1993. In The Nearest Active Galaxies, ed.J Beckman, L Colina, H Netzer, p. 197.\\
Kormendy, J. 2004 in Carnegie Observatories Astrophysics Series, Vol. 1: Coevolution of Black Holes and Galaxies.Cambridge:Cambridge Univ. Press\\
Kormendy J, Richstone DO.1995, Annu. Rev. Astron. Astrophys. 33:581 \\
Lynden-Bell, D \& Rees, M 1971, MNRAS 152, 461\\
Magorrian J, Tremaine S, Richstone D, Bender R, Bower G, \etal 1998, AJ, 115, 2285\\
Mayes, Pearce \& Evans 1985, Astron. Astrophys. 143, 347\\
Murakami, H. \etal 2007, PASJ, 59, 369\\
Prugniel, P., \& Simien, F. 1996, A\&A, 309, 749\\
Riffel, R. \etal 2006, A\&A, 457, 61\\
Sadler, EM, Jenkins, CR, \& Kotanyi, CG 1989, MNRAS, 240, 591\\
Slee, OB, Sadler, E. \& Reynolds, J. 1994, MNRAS, 269, 928\\
Sarzi, M., et al. 2001 ApJ 550, 65\\
Springel V, Di Matteo T, Hernquist L.2005, MNRAS, 361, 776 \\
van der Marel, R., \& van den Bosch, F.C. 1998, AJ, 116, 2220\\
Veron-Cetty, M. \& Veron, P. 2006, A\&A, 455, 773\\ 
Vika, M., et al., 2012, MNRAS, 419, 2264\\
Woo, J. \etal 2012, in preparation

%\end{thebibliography}

%\end{multicols}
% It is preferable to embed your figures in the text as in the following example
\pagebreak
%\vskip 5 truein

\vfill\break

%%Format tables as in the following example
%\begin{table}[h]
%\begin{center}
%\caption{Example Table}\label{tableexample}
%\begin{tabular}{lcc}
%\hline Column 1 & Column 2 & Column 3 \\
%\hline Table Content$^a$ \\
%\hline
%\end{tabular}
%\medskip\\
%$^a$Table footnotes go here.\\
%\end{center}
%\end{table}

\centerline{\bf Table 1: Runs}
\begin{tabbing}
nnn\=Telescopesssssssssss\=Spectrographss\=Datesssssssssssss\=Resolutionss\=Slitsssssssssss\=Wavelength range\kill
\# \>Telescope\>Spectrograph\>Dates\>Resolution\>Slit\>Wavelength range\\
1\>Hale 5m Palomar\>Triplespec\>14--15/9/2011\>~2600\>1$^{\prime\prime}$ 4 pixels\>1.0--2.4$\mu m$\\
2\>Mayall 4m KPNO\>Flamingos\>6--11/11/2011\>~1000\>1$^{\prime\prime}$ 3 pixels\>0.9--1.8$\mu m$\\
3\>Hale 5m Palomar\>Triplespec\>2/1/2012\>~2600\>1$^{\prime\prime}$\>1.0--2.4$\mu m$\\
4\>Mayall 4m KPNO\>Flamingos\>4--9/02/2012\>~1000\>1$^{\prime\prime}$\>0.9--1.8$\mu m$\\
5\>Hale 5m Palomar\>Triplespec\>29--30/5/2012\>~2600\>1$^{\prime\prime}$\>1.0--2.4$\mu m$\\
Note (1) The following radio active early type galaxies have Palomar Triplespec spectra from \\Woo \etal (2012):\\
NGC 3245, 3607, 3608, 4261, 4291, 4374, 4459, 4473, 4486, 4564, 4596, 4649, 4697, 4742, 6251, 7052.\\
None of these galaxies have emission lines in H band.\\
Note (2) The Triplespec detector format is 2048 $\times$ 1024; the Flamingos detector is 2048$^2$ pixels.
\end{tabbing}
\pagebreak

\small
\centerline{\bf Table 2: Emission line nuclei}
\begin{tabbing}
namesssssss\= sssssssssss\= ssssssssssssss\= ssssss\= ssssssssssssssssssss\= ssssssss\= sssssssss\= sssssssssss\= ssssssssssss\= sssss\= \kill
NAME\>  RA\>  DEC \> Run \> Br$\gamma$ Flux \> SFR \> Remarks\> \> P$_{1.4}$ \>redshift\\ %	K	TYPE	S	err_S	Dist	M(K)	P1/4	Spectrum?	M_BH	SoI_pc	SoI_arcsec	Done
\>  ~~~~2000\>  \> \> ergs/cm$^2$/s\> M$_\odot$/yr \> \> \> W/Hz\\%> 10$^8$M$_\odot$ \\ %	K	TYPE	S	err_S	Dist	M(K)	P1/4	Spectrum?	M_BH	SoI_pc	SoI_arcsec	Done
NGC128\>  00:29:15 \>	+02:51:51 \> 1 \> 1.11$\pm$0.04E-15 \> 0.38 \>1.74$\mu$ unident.\> \> 6.0E+20 \>0.014 \\ 	%	8.51	-2	1.5	0.5	58	-25.3	6.00E+020	n			39	0.14	TRUE
NGC524\>  01:24:48 \>	+09:32:19 \> 1 \> 1.85$\pm$0.06E-15\>0.18 \>very broad 1.6$\mu$ \>\> 2.1E+20 \> 0.008\\  %	7.13	-1	3.1	0.4	24	-24.77	2.10E+020	e			30	0.26	TRUE
NGC547\>  01:26:01 \>	--01:20:42 \> 1 \> 3.97$\pm$0.04E-16 \>0.18 \> \> \> 4.0E+24 \>0.018\\		%	8.48	-5	5.9	500	76	-25.92	4.00E+024	x			54	0.15	TRUE
NGC665\>  01:44:56 \>	+10:25:23\> 1 \> 6.92$\pm$0.03E-16 \> 0.38 \> \> \> 6.3E+21 \>0.018\\	%	8.84	-2	9.6	1.1	74	-25.5	6.30E+021	n			43	0.12	TRUE
NGC708\>  01:52:46 \>	+36:09:06\> 1 \> 9.1$\pm$0.9E-16\>0.37\> \>\> 3.5E+22 \>0.016\\	%	8.54	-5	66	2	67	-25.57	3.50E+022	n			45	0.14	TRUE
NGC1052\>  02:41:05 \>	--08:15:21 \> 1 \> \>\>1.26, 1.64$\mu$ [FeII]\> \> 5.1E+22 \>0.005 \\  %	7.44	-5	1.1	100	19	-24	5.10E+022	x		20	0.22	TRUE
NGC1316\>  03:22:42 \>	--37:12:29\> 3 \> 2.81$\pm$0.05E-14 \>1.0\>1.26, 1.64$\mu$ [FeII]\> \> 7.6E+24 \>0.006\\ %	5.58	-2	1.5	10000	21	-26.03	7.60E+24	x	1.10E+09	57	0.56	TRUE
NGC1521\>  04:08:19 \>	--21:03:07 \> 3 \> 4.62$\pm$0.05E-15 \>1.5 \> \> \> 2.3E+21 \>0.014\\ %	8.66	-5	4.2	0.5	67	-25.48	2.30E+21	x	6.90E+08	43	0.13	TRUE
NGC1550\>  04:19:38 \>	+02:24:35\> 1 \> 1.27$\pm$0.03E-15 \> 0.31\> \> \> 5.3E+21 \>0.012 \\   	%	8.72	-3.2	17	2	52	-24.85	5.30E+021	n			31	0.12	TRUE
NGC2110\>  05:52:11 \>	--07:27:22\> 1 \> \> \> \> \> 3.9E+22 \>0.008 \\  					%	8.01	-3	3	10	33	-24.58	3.90E+022	x			27	0.17	TRUE
NGC2128\>  06:04:34 \>	+57:37:40\> 1 \>1.7$\pm$0.1E-15 \> 0.31\> \> \> 5.7E+21 \>0.010 \\   					%	8.78	-3	26	1	43	-24.39	5.70E+021	n			25	0.12	TRUE
NGC2273 \>06:50:08 \>+60:50:45 \> 4\>\>\>[FeII] lines\>\>6.8E+21\>0.006\\
UGC3426\>  06:15:36 \>	+71:02:15\> 1 \> 1.2$\pm$0.02E-13 \> 37.2 \> \> \> 4.1E+23 \>0.0135 \\   %	8.9	-2	1.1	100	55	-24.82	4.10E+023	e		31	0.12	TRUE
NGC2768\>  09:11:38 \>	+60:02:14 \> 1 \> \> \> \> \> 8.7E+20 \>0.004 \\   					%	6.98	-5	14	1	22	-24.77	8.70E+020	e			30	0.28	TRUE
NGC2787\>09:19:18\>+69:12:12\> 1\>1.63$\pm$0.05E-14 \>0.31\>\>\> 7.3E+19 \>0.002\\
NGC2859\>  09:24:19 \>	+34:30:49 \> 1 \> 7.7$\pm$0.4E-15 \> 0.49 \> \> \> 7.0E+19 \>0.006 \\  	%	8.03	-1	1.1	0.5	23	-23.78	7.00E+019	e			18	0.16	TRUE
NGC3065\>  10:01:55 \>	+72:10:13\> 1 \> 1.8$\pm$0.1E-15\>0.18 \> \> \> 3.6E+20 \>0.007\\   					%	8.97	-2	4.1	0.4	27	-23.2	3.60E+020	n			14	0.1	TRUE
IC630\> 10:38:33\>--07:10:14\>3\> 1.99$\pm$0.01E-14\>2.1\> \>\> 7.0E+21 \>0.007\\ %  8.63	-2	67	2	30	-23.73	7.00E+21	e	1.50E+08	18	0.12	TRUE
NGC3516\>11:06:47\>+72:34:07\>3\> \>\>Broad P$\beta$\>\>4.8E+21\>0.009\\
%NGC3557\>11:09:57.6\>--37:32:21\>5\> \>\>+\>\>2E+23\>0.010\\
NGC4111\>12:07:03.1\>+43:03:57\>5\>1.5$\pm$0.1E-15\>0.04\>+\>\>2.5E+20\>0.003\\
NGC5273\>13:42:08.3\>+35:39:15\>5\>\>\>Broad P$\gamma$\>\>1.1E+20\>0.004\\
NGC5322\>13:49:15.3\>+60:11:26\>5\>1.9$\pm$0.8E-15\>0.18\>\>\>9.1E+21\>0.006\\
NGC7052\>  21:18:33 \>	+26:26:49 \> 1 \> 2.49$\pm$0.07E-15 \> 1.03 \> \> \> 7.9E+22 \>0.016 \\	%	8.53	-5	1.6	10	64	-25.51	7.90E+022	n			44	0.14	TRUE
%NGC7426\>  22:56:03 \>	+36:21:41 \> 1 \> 3.75$\pm$0.09E-15 \> 2.01 \> \> \> 3.9E+21 \>0.018 \\	%	8.77	-5	6	0.5	73	-25.56	3.90E+021	n			45	0.13	TRUE
NGC7550\>  23:15:16 \>	+18:57:42 \> 1 \> 1.77$\pm$0.04E-15 \> 0.86 \> \> \> 2.6E+22 \>0.017 \\	%	8.86	-3	45	1	69	-25.35	2.60E+022	n			40	0.12	TRUE
NGC7618\>  23:19:47 \>	+42:51:10 \> 1 \> 1.94$\pm$0.04E-15 \> 0.99 \> \> \> 2.3E+22 \>0.017\\	%	8.96	-5	38	2	71	-25.3	2.30E+022	n			39	0.11	TRUE
NGC7626\>  23:20:43 \>	+08:13:01 \> 1 \> 1.86$\pm$0.06E-15 \> 0.43 \> \> \> 2.0E+23 \>0.011\\  	%	8.01	-5	7.4	50	47	-25.35	2.00E+023	e			40	0.18	TRUE
NGC7743\>  23:44:21 \>	+09:56:03 \> 1 \> 1.90$\pm$0.07E-15 \> 0.09 \> \> \> 3.7E+20\>0.006\\  	%	8.39	-1	7.2	0.5	21	-23.19	3.70E+020	e			14	0.13	TRUE
\end{tabbing}
Note (1) the remark on broad Paschen lines applies to all 3 of NGC 3516, 4111 \& 5273.\\
Note (2) the redshifts in column (10) are literature, not observed, values.
%N128 1.7 Jy at 100 um
%N524 2.0
%N665 2.4
%N708 0.7
%N1052 1.6
%N1316 13
%N2110 5.7
%N2128 2.4
%N2273 1.0
%U3426 ~10
%N2768 1.4
%N2787 1.2
%N2859 0.9
%N3065 2
%N3516 2.3
%IC630 ~15
%N5273 1.6
%N7052 1.5
%N7426 1.0
%N7550 0.5
%N7743 3.4

\pagebreak

\leftline{\bf Table 3: Absorption line nuclei}
\begin{tabbing}
namesssssss\= sssssss\= ssssssssssssss\= sssssssssssss\= sssssss\= ssssssssssssss\kill%= \= sssssss\= sssssssssss\= sssssssssss\kill
NAME\>  Run \>S/N\>NAME\>  Run\>S/N\\
NGC16\> 1\>93 \>NGC3091\>4\>3\\
NGC50\> 2 \>8\>NGC3100\> 3\\
NGC57\> 4\>14\>NGC3158\> 2\>23\\ 				
NGC315\> 1\>50\>NGC3226\> 2\>22\\				%	8.56	-5	3.2	0.4	24	-23.3	2.10E+020	e	1.10E+008	14	0.12	TRUE
NGC383\> 1\>\>NGC3245\> 2\>24\\ 				%	7.85	-2	6.7	0.5	21	-23.75	3.50E+020	e	1.60E+008	18	0.18	TRUE
NGC410\> 1\>22\>NGC3348\>4\>8\\
NGC439\>3\>\>NGC7454\> 2\>10\\  				%	8.83	-5	-0.3	0.5	24	-23.06	<4.1e19		n	8.80E+007	13	0.11	TRUE
%  8.63	-2	67	2	30	-23.73	7.00E+21	e	1.50E+08	18	0.12	TRUE
NGC474\> 1\>49 \>NGC3414\> 2\>30\\ 				%	7.97	-2	4.4	0.4	25	-24.04	3.40E+020	e	2.00E+008	21	0.17	TRUE
NGC499\> 4\>12\>NGC3528 \>4\>20\\
NGC533\> 2\>2\>NGC3607 \>4\>38\\
NGC596\> 4\>20\>NGC3619 \>4\>26\\
NGC680\> 4\>13\>NGC3626 \>4\>14\\
NGC741\> 2\>7\>NGC3665 \>4\>34\\
NGC883\> 2\>7\>NGC3801 \>4\>7\\
NGC936\> 2\>10\>NGC3872 \>4\>17\\
NGC1016\> 2 \>2\>NGC3894 \>4\>20\\
NGC1060\> 2 \>15\>NGC3957 \>4\>11\\
NGC1128\> 1 \>70\>NGC3962 \>4\>31\\
NGC1167\> 1 \>52\>NGC3986 \>4\>14\\
NGC1200\> 2 \>\>NGC4008 \>4\>5\\
NGC1209\> 4\>10\> NGC4036 \>4\>31\\
NGC1326\> 3\>74\>NGC4125 \>4\>31\\
NGC1399\> 3 \>95\>NGC4138 \>4\>28\\
NGC1400\> 2 \>\>NGC4220 \>4\>17\\
NGC1407\> 2 \>8\>NGC4589 \>4\>22\\
NGC1453\> 1 \>89\>NGC4636 \>4\>25\\
NGC1531\> 3 \>15\>NGC4778 \>4\>8\\
IC359\> 2 \>\>NGC4782 \>4\>17\\
UGC3024\> 2\>6 \>NGC4802 \>4\>15\\
NGC1587\> 1\>64 \>NGC4825 \>4\>19\\
NGC1600\> 2 \>8\>NGC4874 \>4\>8\\
NGC1573\> 2\>10 \>NGC4984 \>4\>28\\
NGC1653\> 1 \>70\>NGC5044 \>4\>6\\
NGC1684\> 1 \>68\>NGC5077 \>4\>16\\
NGC1726\> 2 \>10\>NGC5198 \>4\>20\\
NGC2089\> 2\>8 \>NGC5444 \>4\>23\\
NGC2208\> 2\>4 \>NGC5490 \>4\>21\\
NGC2256\> 4\>4 \>NGC5532 \>4\>9\\
NGC2258\> 2\>28 \>NGC5631 \>4\>21\\
NGC2320\> 4\>15 \>NGC5838 \>4\>16\\
NGC2314\> 4 \>9\>IC 459 \>3\>26\\
NGC2340\> 2\>8 \>NGC5846\> 1\>32\\   	%	6.91	-5	21	1	25	-25.07	1.60E+021	e			35	0.29	TRUE
NGC2418\> 4\>20 \>NGC6251\> 1\>53\\  					%	8.99	-5	1.8	100	101	-26.03	2.20E+024	e			57	0.12	TRUE
NGC2493\> 3\>85\>NGC6482\> 1 \>49\\   	%	8.34	-5	1.4	0.5	54	-25.31	4.80E+020	e			39	0.15	TRUE
NGC2513\> 2\>30\>NGC6703\> 1\>67 \\   	%	8.21	-2.5	0.5	0.5	27	-23.92	<1.2e20		e			20	0.15	TRUE
NGC2612\>4\>24\>NGC6869\> 2 \>8\\ 				%	8.64	-2	0.4	0.5	32	-23.89	<1.6e20		n	1.80E+008	19	0.12	TRUE
UGC3789\>4\>15\>NGC6903\> 2\>8\\ 				%	8.31	-2.8	0.7	0.5	45	-24.95	<3.8e20		n	4.40E+008	33	0.15	TRUE
NGC2629\> 3\>71\>NGC7242\> 2\>7 \\ 				%	8.28	-3.5	0.7	0.5	78	-26.18	<1.2e21		n	1.20E+009	61	0.16	TRUE
NGC2685\> 1\>47\>NGC7265\> 2\>13 \\  				%	8.65	-3	0.5	0.5	71	-25.59	<8.4e20		n	7.60E+008	45	0.13	TRUE
NGC2693\> 3\>82\>NGC7391\> 2\>11 \\ 					%	8.59	-5	4.8	0.4	42	-24.51	1.00E+021	n	3.00E+008	26	0.13	TRUE
NGC2749\> 2 \>23\>NGC7436\> 2\>22 \\ 				%	8.97	-5	26	1	101	-26.05	3.20E+022	n	1.10E+009	57	0.12	TRUE 			
NGC2872\> 3 \>78\>NGC7562\> 2\>7 \\  				%	8.28	-5	0.5	0.5	57	-25.51	<5.5e20		n	7.10E+008	44	0.16	TRUE
NGC2911\> 1 \>19\>NGC7600\> 2 \>9\\ 				%	8.89	-3	0.7	0.5	47	-24.48	<4.3e20		n	2.90E+008	26	0.11	TRUE
NGC2974\>4\>55\>NGC7711\> 1\>83\\   					%	8.88	-2	1.1	0.5	56	-24.84	4.10E+020	n			31	0.11	TRUE
NGC3078\> 3 \>103\>NGC7785\> 1 \>69\\  				 	%	8.43	-5	33	2	52	-25.16	1.10E+022	n			37	0.15	TRUE
NGC4551\>5\>50\>NGC5796 \>4\>23\\
NGC5813\>5\>54\>NGC4143\>5\>87\\
NGC4203\>5\>97\>NGC4150\>5\>64\\
NGC4278\>5\>83\>NGC4710\>5\>29\\
NGC4866\>5\>47\>NGC5353\>5\>93\\
NGC3557\>\>NGC7426\>1\\
\end{tabbing}

\pagebreak

%\centerline{\bf Table 4: 
\begin{deluxetable}{lllllllll}
\tablecaption{Table 4: BH masses for Brown \etal (2011) galaxies}
%\tablenumber{4}
\tablehead{\colhead{Name}& 
\colhead{RA}&
\colhead{DEC}&
\colhead{m-M}&
\colhead{~}& 
\colhead{K}&% (2MASS)}&
\colhead{$\sigma$}& 
\colhead{M(BH)$_K$}& 
\colhead{M(BH)$_\sigma$ Notes}}
%\colhead{Notes}}

\startdata
 &J2000&&mag&&2MASS&km s$^{-1}$&10$^8$  M$_\sun$&10$^8$  M$_\sun$\\
%&&[1]&[2]~~~[3]&[4]&[5]&[6]&[7]&[8]\\
 NGC0016  & 00:09:04  & +27:43:45  & 33.2 & CV  & 8.78 & 180  & 3.18  & 0.99\\
 NGC0050  & 00:14:44  & --07:20:42  & 34.4 & CV  & 8.67 & 264 & 11.77  & 6.40\\
 NGC0057  & 00:15:30  & +17:19:42  & 34.4 & CV  & 8.68 & 326 & 11.41 & 17.84\\
 NGC0128  & 00:29:15  & +02:51:50  & 33.9 & CV  & 8.52 & 215  & 7.90  & 2.36\\
 NGC0315  & 00:57:48  & +30:21:29  & 34.2 & CV  & 7.95 & 279 & 18.89  & 8.37\\
 NGC0383  & 01:07:24  & +32:24:44  & 34.2 & CV  & 8.48 & 279 & 12.10  & 8.37\\
 NGC0410  & 01:10:58  & +33:09:08  & 34.3 & CV  & 8.38 & 300 & 14.41 & 11.91\\
 NGC0439  & 01:13:47  & --31:44:49  & 34.5 & CV  & 8.70 & 221 & 12.65  & 2.70\\
 NGC0474  & 01:20:06  & +03:24:55  & 32.5 & CV  & 8.55 & 163  & 2.18  & 0.61\\
 NGC0499  & 01:23:11  & +33:27:36  & 33.9 & CV  & 8.73 & 266  & 6.94  & 6.64\\
 NGC0524  & 01:24:47  & +09:32:19  & 32.6 & CV  & 7.16 & 253  & 8.77  & 5.20\\
 NGC0533  & 01:25:31  & +01:45:32  & 34.4 & CV  & 8.44 & 279 & 14.92  & 8.37\\
 NGC0547  & 01:26:00  & --01:20:42  & 34.4 & CV  & 8.49 & 260 & 13.85  & 5.94\\
 NGC0596  & 01:32:51  & --07:01:53  & 32.1 & CV  & 7.97 & 152  & 2.40  & 0.44\\
 NGC0665  & 01:44:56  & +10:25:23  & 34.4 & CV  & 8.88 & 190  & 9.37  & 1.29\\
 NGC0680  & 01:49:47  & +21:58:14  & 33.0 & CV  & 8.73 & 212  & 2.70  & 2.20\\
 NGC0708  & 01:52:46  & +36:09:07  & 34.1 & CV  & 8.57 & 230  & 9.94  & 3.27\\
 NGC0741  & 01:56:20  & +05:37:43  & 34.4 & CV  & 8.29 & 291 & 17.40 & 10.27\\
 NGC0883  & 02:19:05  & --06:47:27  & 34.3 & CV  & 8.89 & 287  & 8.84  & 9.60\\
 NGC0936  & 02:27:37  & --01:09:21  & 31.4 & CV  & 6.91 & 189  & 3.34  & 1.26\\
 NGC1016  & 02:38:19  & +02:07:09  & 34.1 & $D_n-\sigma$  & 8.58 & 302 & 10.02 & 12.30\\
 NGC1052  & 02:41:04  & --08:15:20  & 31.4 & SBF  & 7.45 & 207  & 2.16  & 1.96\\
 NGC1060  & 02:43:15  & +32:25:29  & 34.3 & $D_n-\sigma$  & 8.20 & 303 & 16.18 & 12.50\\
 NGC1128  & 02:57:42  & +06:01:29  & 34.9 & CV  & 8.98 & \nd & 14.72  & \nd \\%& \\
 NGC1167  & 03:01:42  & +35:12:20  & 34.2 & CV  & 8.64 & 171  & 9.68  & 0.78\\
 NGC1200  & 03:03:54  & --11:59:29  & 33.7 & CV  & 8.58 & 204  & 6.56  & 1.83\\
 NGC1209  & 03:06:02  & --15:36:40  & 32.8 & SBF  & 8.32 & 230  & 3.37  & 3.27\\
 NGC1316  & 03:22:41  & --37:12:29  & 31.7 & SBF  & 5.59 & 228 & 16.23  & 3.14\\
 NGC1326  & 03:23:56  & --36:27:52  & 31.4 & TF  & 7.45 & 118  & 2.04  & 0.13\\
 NGC1399  & 03:38:29  & --35:27:02  & 31.6 & SBF  & 6.31 & 346  & 7.64 & 23.8 ~5.1 $\pm$ 0.7 (3)\\%$M_{\odot} = 
 NGC1400  & 03:39:30  & --18:41:17  & 31.1 & SBF  & 7.81 & 252  & 1.06  & 5.10\\
 NGC1407  & 03:40:11  & --18:34:48  & 31.1 & SBF  & 6.70 & 271  & 3.12  & 7.27\\
 NGC1453  & 03:46:27  & --03:58:08  & 33.7 & $D_n-\sigma$  & 8.12 & 328 & 10.38 & 18.38\\
 NGC1521  & 04:08:18  & --21:03:06  & 34.1 & SBF  & 8.68 & 242  & 8.98  & 4.19\\
 NGC1531  & 04:11:59  & -32:51:03  & 30.6 & TF  & 8.42 & 112  & 0.36  & 0.10\\
  IC0359  & 04:12:28  & +27:42:06  & 33.8 & CV  & 8.47 & \nd  & 7.54  & \nd\\
 NGC1550  & 04:19:37  & +02:24:35  & 33.6 & CV  & 8.77 & 308  & 4.68 & 13.54\\
UGC3024  & 04:22:26  & +27:17:52  & 34.3 & CV  & 9.25 & \nd  & 6.10  & \nd\\
 NGC1587  & 04:30:39  & +00:39:42  & 32.8 & $D_n-\sigma$  & 8.51 & 227  & 2.86  & 3.07\\
 NGC1600  & 04:31:39  & --05:05:09  & 33.5 & $D_n-\sigma$  & 8.04 & 334  & 9.41 & 20.07\\
 NGC1573  & 04:35:03  & +73:15:44  & 33.9 & $D_n-\sigma$  & 8.55 & 303  & 7.80 & 12.50\\
 NGC1653  & 04:45:47  & --02:23:33  & 33.8 & SBF  & 8.97 & 246  & 4.95  & 4.54\\
 NGC1684  & 04:52:31  & --03:06:21  & 33.9 & CV  & 8.69 & 306  & 7.25 & 13.11\\
 NGC1726  & 04:59:41  & --07:45:19  & 33.0 & $D_n-\sigma$  & 8.61 & 246  & 3.20  & 4.54\\
 NGC2089  & 05:47:51  & --17:36:09  & 33.0 & CV  & 8.81 & 206  & 2.57  & 1.92\\
 NGC2110  & 05:52:11  & --07:27:22  & 32.6 & CV  & 8.14 & 255  & 3.46  & 5.41\\
 NGC2128  & 06:04:34  & +57:37:40  & 33.1 & CV  & 8.83 & 261  & 2.87  & 6.05\\
UGC3426  & 06:15:36  & +71:02:15  & 33.8 & CV  & 8.97 & 283  & 4.65  & 8.97\\
 NGC2208  & 06:22:34  & +51:54:34  & 34.5 & CV  & 9.04 & 225  & 9.34  & 2.94\\
 NGC2256  & 06:47:13  & +74:14:11  & 34.3 & CV  & 8.67 & 221 & 10.73  & 2.70\\
 NGC2258  & 06:47:46  & +74:28:54  & 33.6 & SBF  & 8.23 & 287  & 7.93  & 9.60\\
 NGC2273  & 06:50:08  & +60:50:44  & 32.3 & TF  & 8.48 & 123  & 1.77  & 0.16\\
 NGC2273  & 06:50:08  & +60:50:44  & 32.3 & TF  & 8.48 & 123  & 1.77  & 0.16\\
 NGC2320  & 07:05:41  & +50:34:51  & 34.5 & SNIa  & 8.85 & 315 & 10.69 & 15.10\\
 NGC2314  & 07:10:32  & +75:19:36  & 34.0 & $D_n-\sigma$  & 8.88 & 286  & 6.70  & 9.44\\
  IC0459  & 07:10:38  & +50:10:37  & 34.5 & CV  & 11.29 & \nd  & 0.99  & \nd\\
 NGC2340  & 07:11:10  & +50:10:28  & 34.7 & $D_n-\sigma$  & 8.88 & 246 & 13.18  & 4.54\\
UGC3789  & 07:19:30  & +59:21:18  & 33.3 & CV  & 9.51 & \nd  & 1.75  & \nd\\
 NGC2418  & 07:36:37  & +17:53:02  & 34.2 & CV  & 8.95 & 247  & 7.50  & 4.63\\
 NGC2493  & 08:00:23  & +39:49:49  & 33.7 & CV  & 8.83 & 249  & 4.94  & 4.82\\
 NGC2513  & 08:02:24  & +09:24:47  & 33.9 & $D_n-\sigma$  & 8.74 & 274  & 6.51  & 7.67\\
 NGC2612  & 08:33:50  & --13:10:28  & 31.8 & CV  & 8.78 & \nd  & 0.83  & \nd\\
 NGC2629  & 08:47:15  & +72:59:08  & 33.6 & CV  & 8.85 & 298  & 4.37 & 11.53\\
 NGC2685  & 08:55:34  & +58:44:03  & 30.9 & TF  & 8.35  & 94  & 0.54  & 0.04\\
 NGC2693  & 08:56:59  & +51:20:50  & 34.0 & $D_n-\sigma$  & 8.60 & 349  & 8.27 & 24.84\\
 NGC2749  & 09:05:21  & +18:18:47  & 33.9 & $D_n-\sigma$  & 8.93 & 260  & 5.50  & 5.94\\
 NGC2768  & 09:11:37  & +60:02:14  & 31.8 & SBF  & 7.00 & 181  & 4.53  & 1.02\\
 NGC2787  & 09:19:18  & +69:12:11  & 29.4 & SBF  & 7.26 & 194  & 0.35  & 1.43 ~~1.04$^{+0.36}_{-0.64}$  (1)\\%\\%& $M_{\odot} = 
 NGC2859  & 09:24:18  & +34:30:48  & 32.0 & TF  & 8.04 & 181  & 2.17  & 1.02\\
 NGC2872  & 09:25:42  & +11:25:55  & 33.9 & $D_n-\sigma$  & 8.72 & 284  & 6.75  & 9.12\\
 NGC2911  & 09:33:46  & +10:09:07  & 33.3 & CV  & 8.71 & 217  & 3.72  & 2.47\\
 NGC2974  & 09:42:33  & --03:41:57  & 31.7 & SBF  & 6.25 & 238  & 8.51  & 3.87\\
 NGC3078  & 09:58:24  & --26:55:33  & 32.7 & SBF  & 7.88 & 251  & 4.97  & 5.01\\
 NGC3091  & 10:00:14  & --19:38:10  & 33.6 & $D_n-\sigma$  & 8.09 & 321  & 9.20 & 16.55\\
 NGC3100  & 10:00:40  & --31:39:51  & 32.8 & CV  & 8.08 & 200  & 4.27  & 1.66\\
 NGC3065  & 10:01:55  & +72:10:13  & 32.5 & TF  & 8.99 & 160  & 1.34  & 0.56\\
 NGC3158  & 10:13:50  & +38:45:53  & 34.5 & $D_n-\sigma$  & 8.80 & 343 & 11.47 & 22.84\\
 NGC3226  & 10:23:26  & +19:53:53  & 31.9 & SBF  & 8.57 & 193  & 1.10  & 1.40\\
 NGC3245  & 10:27:18  & +28:30:26  & 31.6 & SBF  & 7.86 & 210  & 1.70  & 2.10 ~~2.04 $\pm 0.49$ (4)\\%\\%& $M_{\odot} = 
  IC0630  & 10:38:33  & --07:10:14  & 32.4 & CV  & 8.65 & \nd  & 1.73  & \nd\\
 NGC3348  & 10:47:09  & +72:50:22  & 32.7 & $D_n-\sigma$  & 7.96 & 238  & 4.45  & 3.87\\
 NGC3414  & 10:51:16  & +27:58:28  & 32.0 & SBF  & 7.98 & 237  & 2.25  & 3.79\\
 NGC3516  & 11:06:47  & +72:34:06  & 33.0 & TF  & 8.51 & 157  & 3.34  & 0.51\\
 NGC3497  & 11:07:18  & --19:28:17  & 33.5 & CV  & 8.67 & 224  & 5.02  & 2.88 ==NGC3528\\
 NGC3607  & 11:16:54  & +18:03:09  & 31.8 & SBF  & 6.99 & 224  & 4.74  & 2.88 ~~1.33 $\pm 0.44$ (5)\\%$\\%& $M_{\odot} = 
 NGC3619  & 11:19:21  & +57:45:28  & 32.2 & TF  & 8.57 & 162  & 1.57  & 0.60\\
 NGC3626  & 11:20:03  & +18:21:26  & 31.5 & SBF  & 8.16 & 142  & 1.17  & 0.31\\
 NGC3665  & 11:24:43  & +38:45:46  & 32.6 & TF  & 7.68 & 184  & 5.11  & 1.11\\
 NGC3801  & 11:40:16  & +17:43:41  & 33.4 & CV  & 8.88 & 198  & 3.67  & 1.58\\
 NGC3872  & 11:45:49  & +13:46:00  & 33.3 & $D_n-\sigma$  & 8.51 & 260  & 4.75  & 5.94\\
 NGC3894  & 11:48:50  & +59:24:56  & 33.2 & CV  & 8.56 & 265  & 4.28  & 6.52\\
 NGC3957  & 11:54:01  & --19:34:06  & 32.2 & TF  & 8.69 & \nd  & 1.36  & \nd\\
 NGC3962  & 11:54:40  & --13:58:29  & 32.7 & SBF  & 7.67 & 233  & 6.17  & 3.49\\
 NGC3986  & 11:56:44  & +32:01:18  & 33.3 & CV  & 8.98 & 196  & 2.89  & 1.50\\
 NGC4008  & 11:58:17  & +28:11:32  & 33.4 & $D_n-\sigma$  & 8.83 & 226  & 3.65  & 3.01\\
 NGC4036  & 12:01:26  & +61:53:44  & 31.9 & TF  & 7.56 & 181  & 3.18  & 1.02\\
 NGC4125  & 12:08:06  & +65:10:26  & 31.9 & SBF  & 6.86 & 227  & 5.95  & 3.07\\
 NGC4138  & 12:09:29  & +43:41:06  & 30.7 & SBF  & 8.20 & 140  & 0.51  & 0.29\\
 NGC4220  & 12:16:11  & +47:52:59  & 31.4 & TF  & 8.13 & 125  & 1.05  & 0.17\\
 NGC4589  & 12:37:25  & +74:11:30  & 31.7 & SBF  & 7.76 & 224  & 2.09  & 2.88\\
 NGC4636  & 12:42:49  & +02:41:15  & 30.8 & SBF  & 6.42 & 203  & 3.24  & 1.78\\
 NGC4778  & 12:53:06  & --09:12:14  & 33.9 & CV  & 8.63 & 232  & 7.35  & 3.42   ==NGC4759\\
 NGC4782  & 12:54:35  & --12:34:06  & 33.9 & $D_n-\sigma$  & 7.75 & 326 & 16.64 & 17.84\\
 NGC4802  & 12:55:49  & --12:03:18  & 30.3 & SBF  & 8.50 & \nd  & 0.26  & \nd\\
 NGC4825  & 12:57:12  & --13:39:52  & 33.9 & CV  & 8.51 & 308  & 8.51 & 13.54\\
 NGC4874  & 12:59:35  & +27:57:34  & 35.0 & GCLF  & 8.86 & 279 & 17.49  & 8.37\\
 NGC4984  & 13:08:57  & --15:30:58  & 31.6 & TF  & 7.74 & \nd  & 1.98  & \nd\\
 NGC5044  & 13:15:23  & --16:23:04  & 32.5 & SBF  & 7.71 & 242  & 4.56  & 4.19\\
 NGC5077  & 13:19:31  & --12:39:21  & 32.9 & $D_n-\sigma$  & 8.22 & 256  & 4.24  & 5.51\\
 NGC5198  & 13:30:11  & +46:40:14  & 33.4 & $D_n-\sigma$  & 8.90 & 196  & 3.49  & 1.50\\
 NGC5444  & 14:03:24  & +35:07:55  & 33.7 & $D_n-\sigma$  & 8.84 & 229  & 5.13  & 3.21\\
 NGC5490  & 14:09:57  & +17:32:43  & 34.5 & SNIa  & 8.92 & 288 & 10.59  & 9.77\\
 NGC5532  & 14:16:52  & +10:48:32  & 35.1 & CV  & 8.76 & 294 & 20.22 & 10.80\\
 NGC5631  & 14:26:33  & +56:34:57  & 32.2 & SBF  & 8.47 & 171  & 1.72  & 0.78\\
 NGC5796  & 14:59:23  & --16:37:25  & 32.9 & $D_n-\sigma$  & 8.15 & 273  & 4.36  & 7.53\\
 NGC5838  & 15:05:26  & +02:05:58  & 32.3 & TF  & 7.58 & 266  & 4.26  & 6.64\\
 NGC5846  & 15:06:29  & +01:36:20  & 32.0 & SBF  & 6.93 & 239  & 6.01  & 3.95\\
 NGC6251  & 16:32:31  & +82:32:16  & 35.1 & CV  & 9.03 & 325 & 15.63 & 17.57 ~~6.0 $\pm 0.2 (6)$\\%\\%& $M_{\odot} = 
 NGC6482  & 17:51:48  & +23:04:18  & 33.7 & CV  & 8.37 & 308  & 7.82 & 13.54\\
 NGC6703  & 18:47:18  & +45:33:01  & 32.1 & SBF  & 8.25 & 178  & 1.95  & 0.94\\
 NGC6869  & 20:00:42  & +66:13:39  & 32.5 & SBF  & 8.71 & 166  & 1.83  & 0.67\\
 NGC7052  & 21:18:33  & +26:26:48  & 34.1 & CV  & 8.57 & 284  & 9.20  & 9.12 ~~3.7$^{+2.6}_{-1.5}$ (2)\\%\\%& $M_{\odot} = 
 NGC7242  & 22:15:39  & +37:17:55  & 34.5 & CV  & 8.33 & \nd & 17.95  & \nd\\
 NGC7265  & 22:22:27  & +36:12:35  & 34.2 & CV  & 8.69 & 258  & 9.79  & 5.72\\
 NGC7391  & 22:50:36  & --01:32:41  & 34.0 & $D_n-\sigma$  & 8.63 & 244  & 8.08  & 4.36\\
 NGC7426  & 22:56:02  & +36:21:40  & 34.3 & CV  & 8.81 & \nd  & 9.55  & \nd\\
 NGC7436  & 22:57:57  & +26:09:00  & 35.0 & CV  & 9.01 & 352 & 15.77 & 25.90\\
 NGC7454  & 23:01:06  & +16:23:18  & 31.9 & SBF  & 8.86 & 114  & 0.86  & 0.11\\
 NGC7454  & 23:01:06  & +16:23:18  & 31.9 & SBF  & 8.86 & 114  & 0.86  & 0.11\\
 NGC7550  & 23:15:16  & +18:57:40  & 34.2 & CV  & 8.91 & 255  & 7.90  & 5.41\\
 NGC7562  & 23:15:57  & +06:41:14  & 33.8 & SBF  & 8.32 & 256  & 9.03  & 5.51\\
 NGC7600  & 23:18:53  & --07:34:49  & 33.4 & CV  & 8.91 & 210  & 3.56  & 2.10\\
 NGC7618  & 23:19:47  & +42:51:09  & 34.3 & CV  & 9.04 & 298  & 7.35 & 11.53\\
 NGC7626  & 23:20:42  & +08:13:01  & 33.7 & SBF  & 8.03 & 271 & 10.66  & 7.27\\
 NGC7711  & 23:35:39  & +15:18:07  & 33.8 & CV  & 8.91 & 180  & 4.93  & 0.99\\
 NGC7743  & 23:44:21  & +09:56:02  & 31.6 & SBF  & 8.42  & 84  & 0.97  & 0.02\\
 NGC7785  & 23:55:19  & +05:54:57  & 33.8 & $D_n-\sigma$  & 8.45 & 255  & 7.98  & 5.41\\
\enddata
%Notes:
\tablenotetext{[1]}{ J2000 coordinates, from the NASA/IPAC Extragalactic Database (NED, http://ned.ipac.caltech.edu)}
\tablenotetext{[2]}{ Distance modulus estimate, using the method identified in column 5, as listed in the NASA/IPAC Extragalactic Database (NED, http://ned.ipac.caltech.edu)}
\tablenotetext{[3]}{ Method used in determining the distance modulus. For uniformity, preference was given to Surface Brightness Fluctuation (SBF) distances, followed by Type Ia Supernovae (SNIa) distances, distances based on the Tully-Fisher (TF) relation and D$_n$-$\sigma$ relations, and redshift distances (CV). The latter are calculated from the recessional velocity (as listed in NED) divided by H$_0$ = 72 km s$^{-1}$ Mpc$^{-1}$ (Freedman et al. 2000). Specific references can be found in NED.}
\tablenotetext{[4]}{ Total K-band magnitude derived from fit extrapolation, as listed in the 2MASS database (http://irsa.ipac.caltech.edu/cgi-bin/Gator/nph-dd, Jarrett 2000)}
\tablenotetext{[5]}{ Average value of the stellar velocity dispersion, computed from all available sources following the precepts of Prugniel and Simien (1996), as listed in the HyperLeda database (http://leda.univ-lyon1.fr/).} 
\tablenotetext{[6]}{ Black hole mass estimated from the K-band magnitude listed in column 6, following equation 7 of Vika et al. (2012) }
\tablenotetext{[7]}{ Black hole mass estimated from the stellar velocity dispersion listed in column 7, following equation 20 of Ferrarese \& Ford (2005)}
\tablenotetext{[8]}{ Black hole masses based on dynamical models applied to spatially resolved kinematics are listed when available. References are: (1) Sarzi et al. (2001); (2) van der Marel \& van den Bosch (1998); (3) Gebhardt et al. (2007); (4) Barth et al. (2001); (5) Gultekin et al. (2009); (6) Ferrarese \& Ford (1999)}
\end{deluxetable}

\begin{figure}[h]
\begin{center}
\includegraphics[scale=.7, angle=-90]{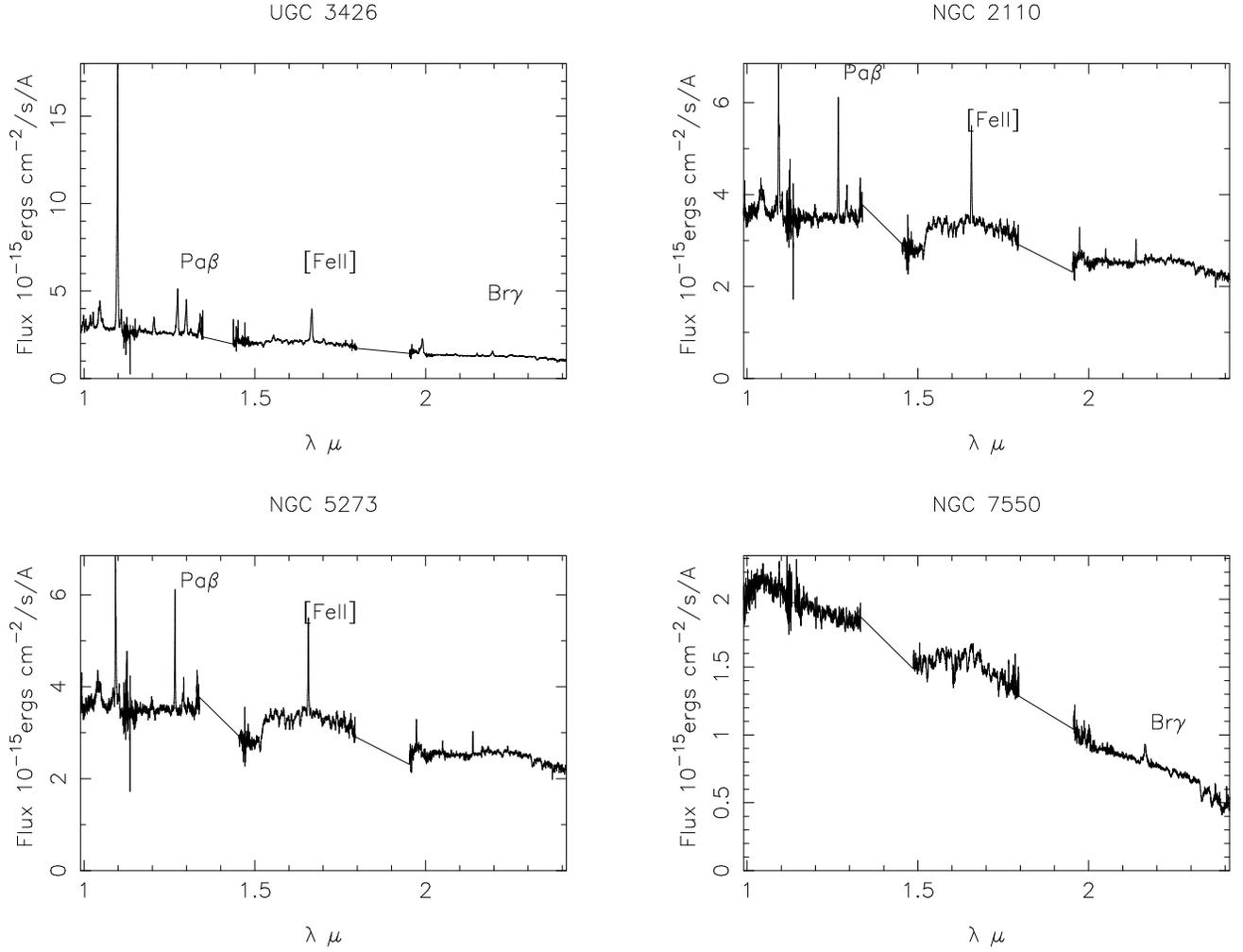}
\caption{Palomar Triplespec spectra (1$^{\prime\prime}$ aperture) of 3 of the strongest emission line 
galaxies encountered in the sample so far. [FeII] is the strong line at 1.64$\mu m$. P$\beta$ is at 1.28 and Br$\gamma$ at 2.18$\mu m$. %NGC 410 is highly reddened ($>$4 mag at 1$\mu m$). The alternative notion, warm dust, is denied by normal stellar CO bandheads in absorption at 2.3$\mu m$. 
NGC 7550 has Br$\gamma$ in emission, indicating a high star formation rate.}
\end{center}
\end{figure}

\begin{figure}[h]
\begin{center}
\includegraphics[scale=.7, angle=-90]{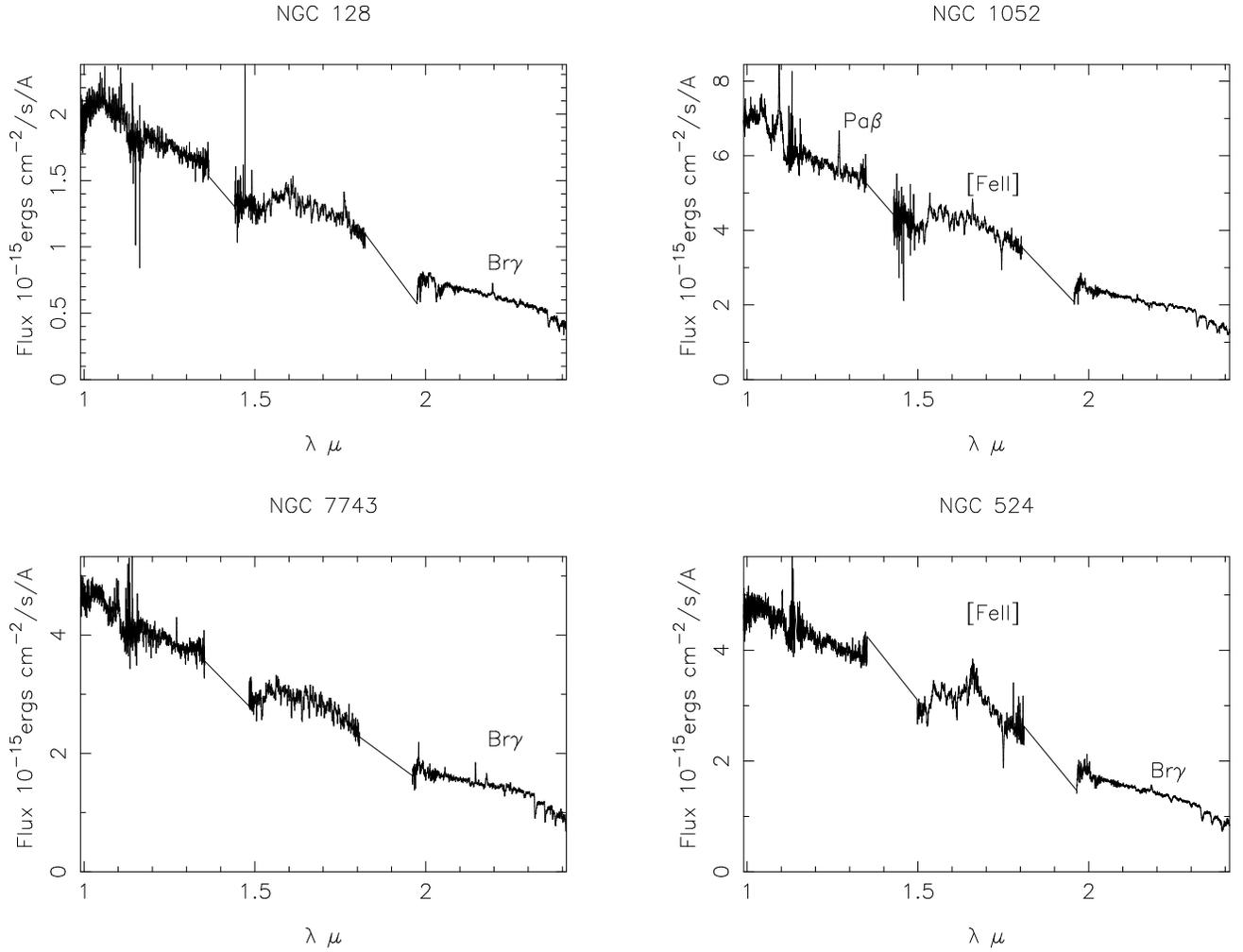}
\caption{Palomar Triplespec spectra of emission line nuclei.
The absorption features beyond 2.3$\mu$ are CO bandheads
from the stellar nucleus.}
\end{center}
\end{figure}

\begin{figure}[h]
\begin{center}
\includegraphics[scale=.45, angle=-90]{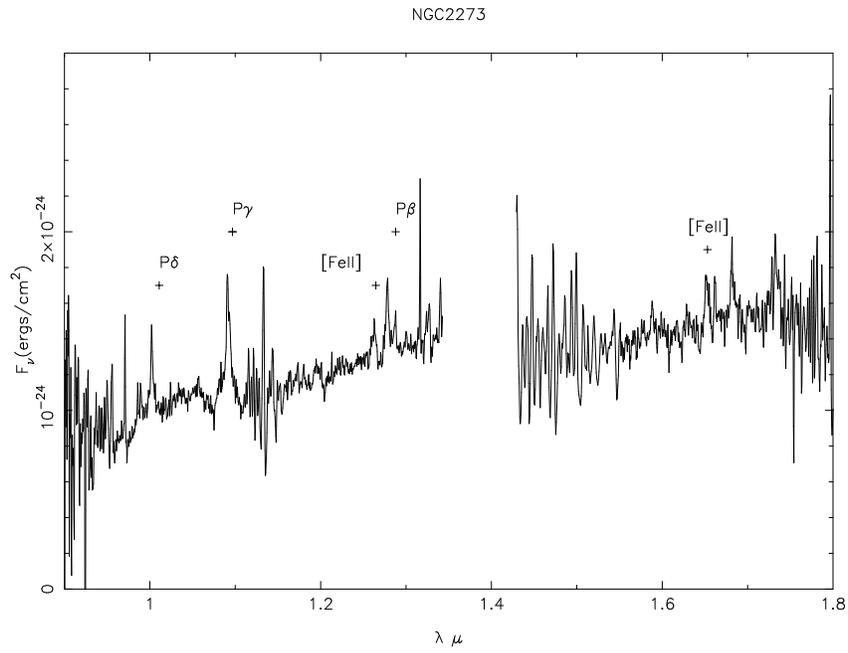}
\caption{Flamingos spectrum of NGC 2273.
%The vertical axis is spectrograph counts after normalised division by an A0V star
%to remove telluric features.
The Paschen emission lines, such as P$\gamma$ 1.09$\mu$, may thus be exaggerated. Three point boxcar smoothing has been applied.
}
\end{center}
\end{figure}

\begin{figure}[h]
\begin{center}
\includegraphics[scale=.7, angle=-90]{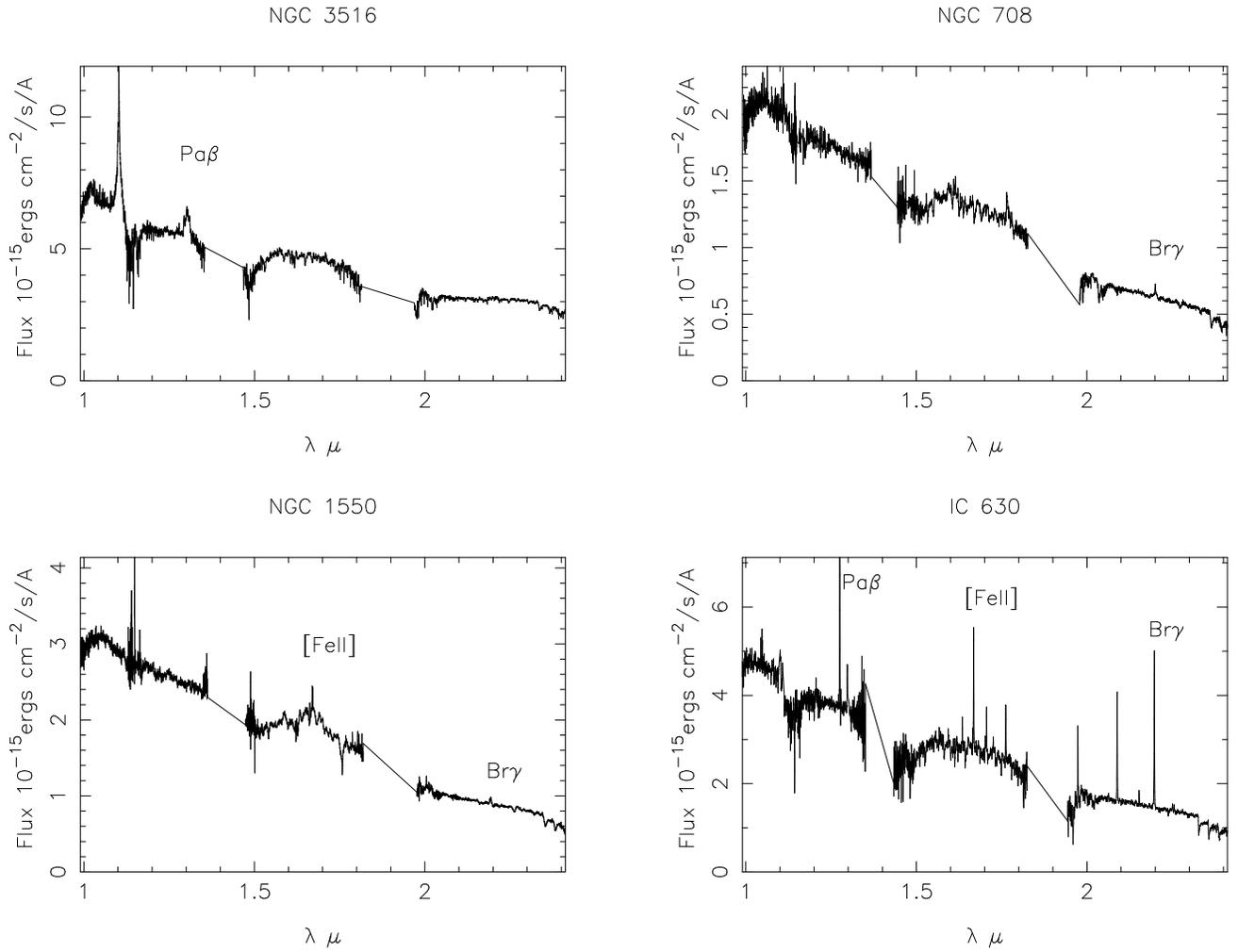}
\caption{Further Palomar spectra of emission line nuclei.}
\end{center}
\end{figure}

\begin{figure}[h]
\begin{center}
\includegraphics[scale=.7, angle=-90]{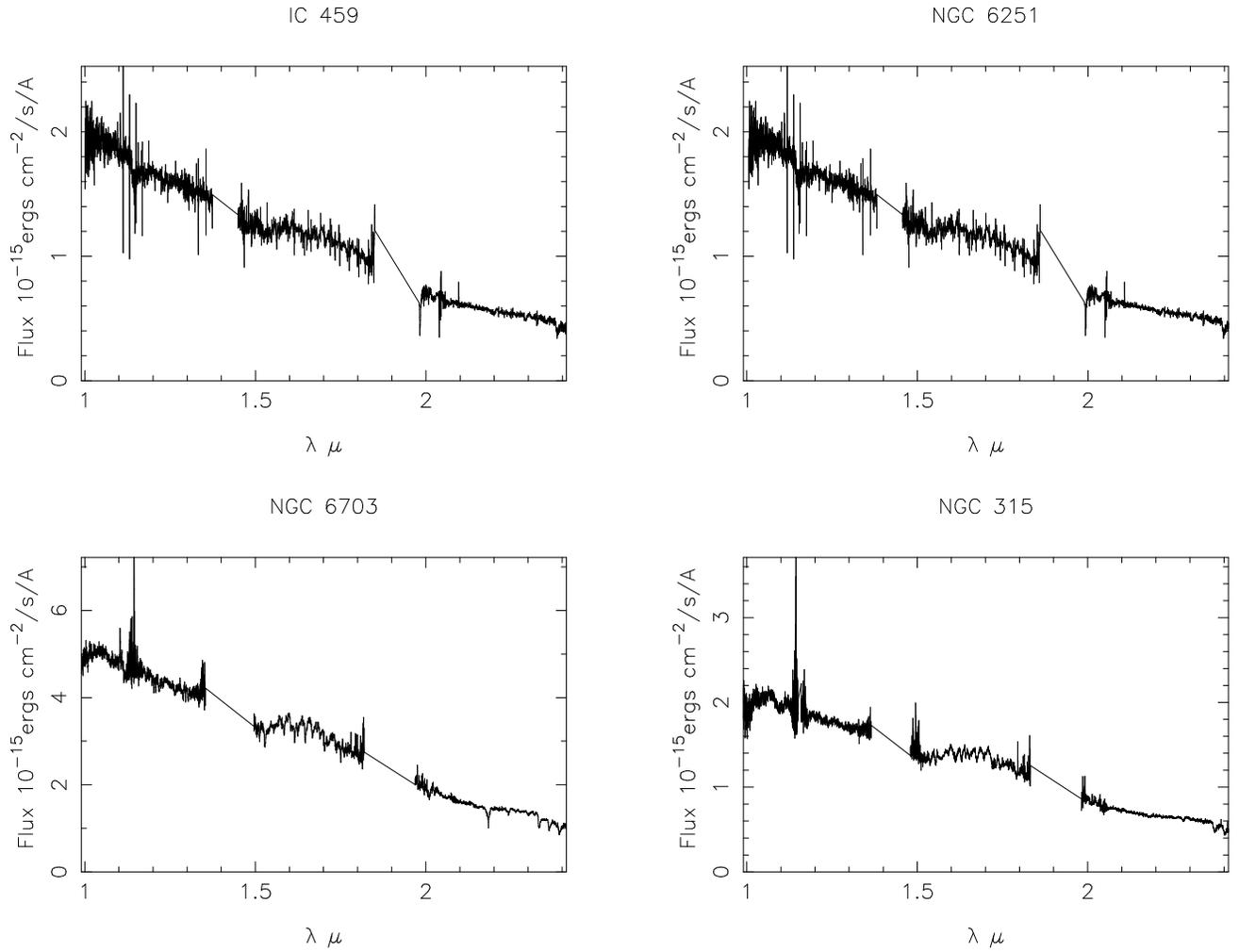}
\caption{Palomar spectra of absorption line nuclei. CO bandheads are seen at the red end of the 2$\mu$ window. NG 6703 has Br$\gamma$ in absorption.}
\end{center}
\end{figure}

\begin{figure}[h]
\begin{center}
\includegraphics[scale=.7, angle=-90]{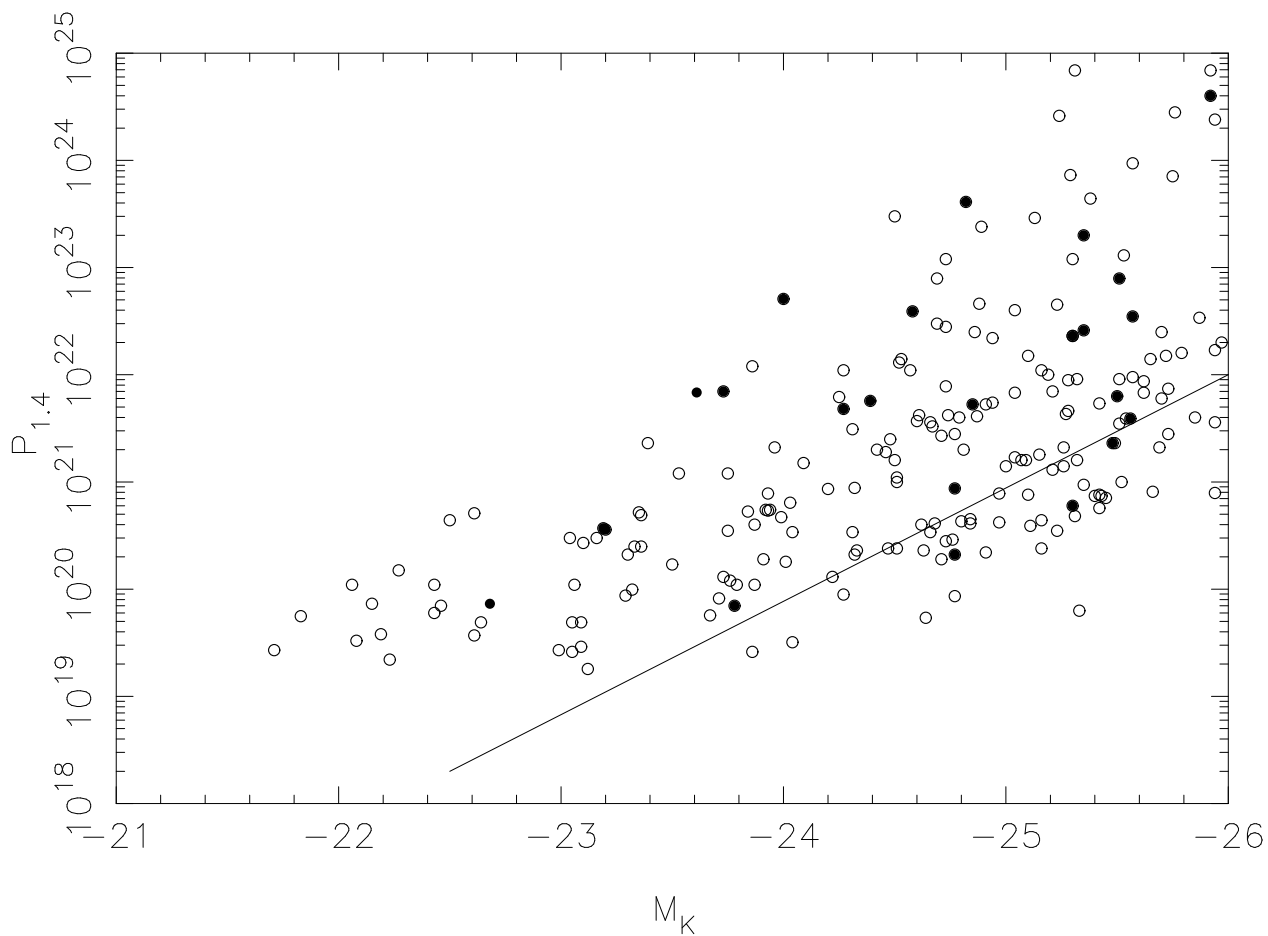}
\caption{Radio power (W/Hz) at 1.4 GHz versus absolute K magnitude,
The data are from Brown \etal (2011). Galaxies whose radio powers are upper limits are not distinguished. Emission line nuclei are
denoted by solid symbols. The line is from the original reference and is the median radio power as a function of K-band luminosity for all ellipticals (including those not detected by NVSS). As this sample only includes radio sources, it is automatically biased towards higher radio powers and is systematically above the line.}
\end{center}
\end{figure}

\begin{figure}[h]
\begin{center}
\includegraphics[scale=.7, angle=-90]{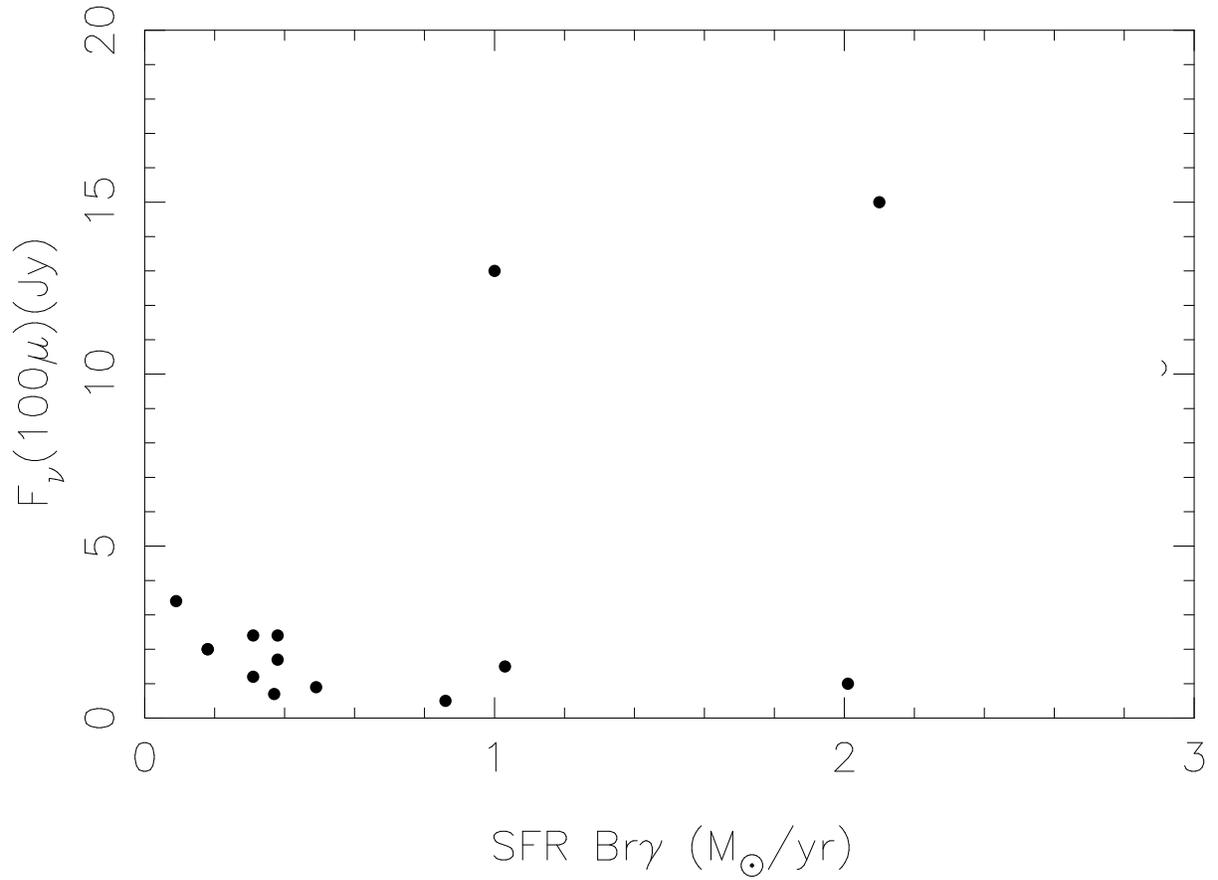}
\caption{Comparison of the Br$\gamma$ star formation rates
with IRAS far infrared fluxes. UGC 3426 is the upper limit at 10 Jy.}
%profiles of two galaxies.
%The spectra have not been corrected to the rest frame.}
\end{center}
\end{figure}

\begin{figure}[h]
\begin{center}
\includegraphics[scale=.7, angle=-90]{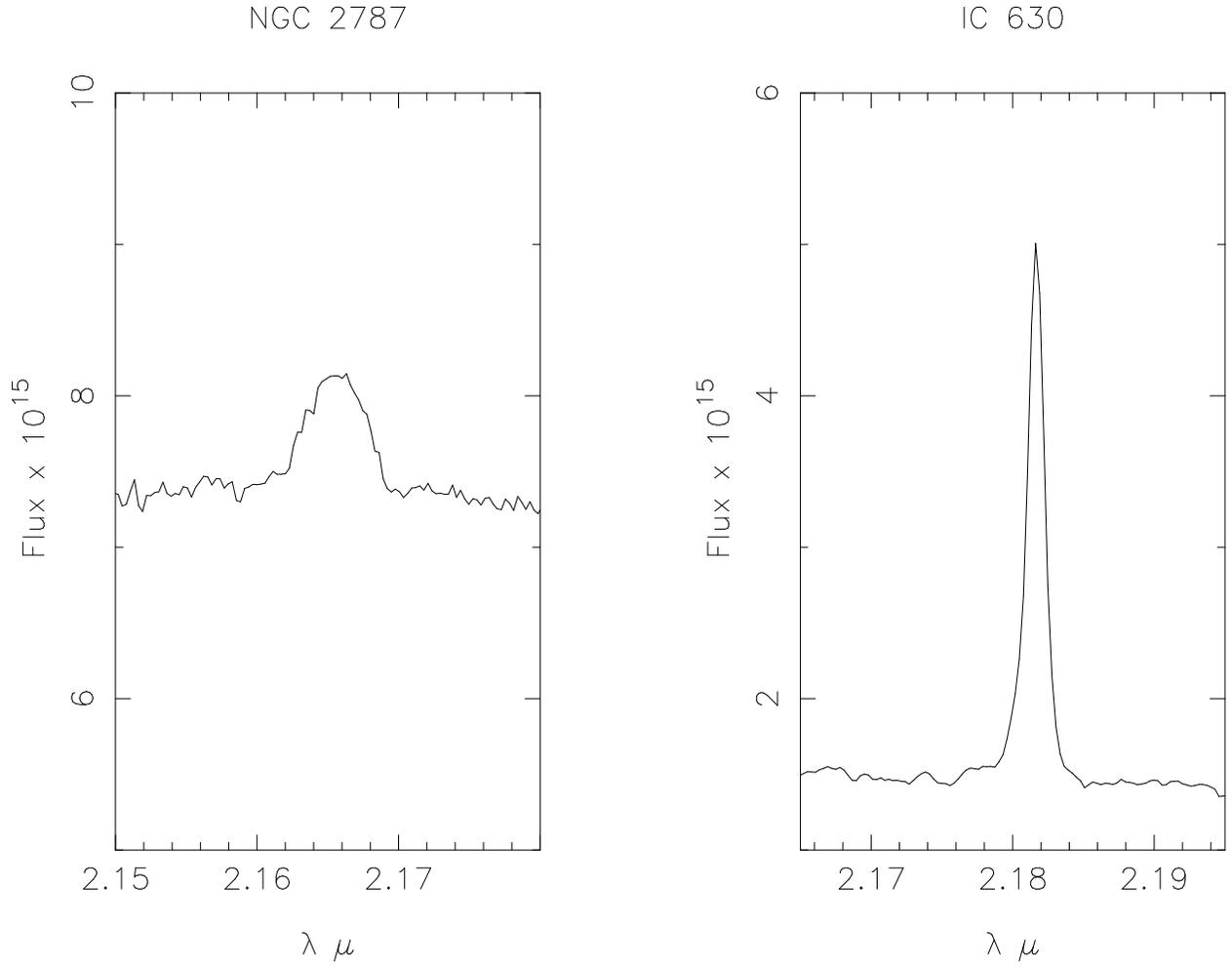}
\caption{Comparison of the Br$\gamma$ profiles of two galaxies.
The spectra have not been corrected to the rest frame.}
\end{center}
\end{figure}

\begin{figure}[h]
\begin{center}
\includegraphics[scale=.7, angle=-90]{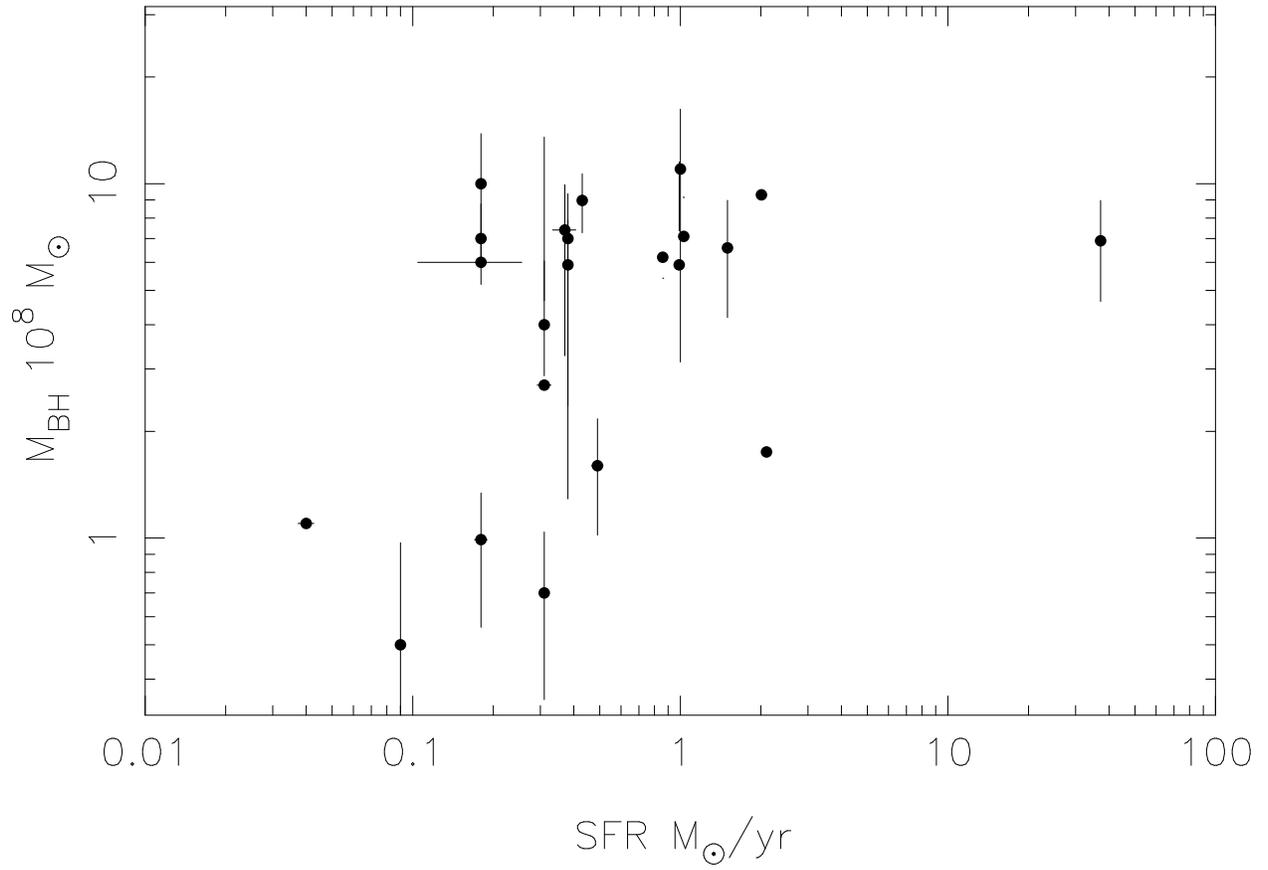}
\caption{SFR and black hole mass from Tables 2 and 4.}
\end{center}
\end{figure}

\end{document}